\documentclass[aps,pra,twocolumn,showpacs,floatfix,10pt,superscriptaddress]{revtex4-1}
\pdfoutput=1

\usepackage{amsmath}
\usepackage{amssymb}
\usepackage{graphicx}

\usepackage{color}

\begin{document}
\title{Spin liquid phases of alkaline-earth-metal atoms at finite temperature}
\author{P. Sinkovicz}
\affiliation{Institute for Solid State Physics and Optics, Wigner Research Centre, Hungarian Academy of Sciences, H-1525 Budapest P.O. Box 49, Hungary}
\author{A. Zamora}
\affiliation{ICFO-Institut de Ci\`encies Fot\`oniques, Mediterranean Technology Park, 08860 Castelldefels (Barcelona), Spain}
\author{E. Szirmai}
\affiliation{BME-MTA Exotic Quantum Phases Research Group, Institute of Physics, Budapest University of Technology and Economics, Budafoki \'ut 8., H-1111 Budapest, Hungary}
\author{M. Lewenstein}
\affiliation{ICFO-Institut de Ci\`encies Fot\`oniques, Mediterranean Technology Park, 08860 Castelldefels (Barcelona), Spain}
\affiliation{ICREA-Instituci\'o Catalana de Recerca i Estudis Avan\c cats, Lluis Companys 23, 08010 Barcelona, Spain}
\author{G. Szirmai}
\affiliation{Institute for Solid State Physics and Optics, Wigner Research Centre, Hungarian Academy of Sciences, H-1525 Budapest P.O. Box 49, Hungary}

\begin{abstract}
We study spin liquid phases of spin-5/2 alkaline-earth-metal atoms on a honeycomb lattice at finite temperatures. Our analysis is based on a Gutzwiller projection variational approach recast to a path-integral formalism. In the framework of a saddle-point approximation we determine spin liquid phases with lowest free energy and study their temperature dependence. We identify a critical temperature, where all the spin liquid phases melt and the system goes to the paramagnetic phase. We also study the stability of the saddle-point solutions and show that a time-reversal symmetry breaking state, a so called chiral spin liquid phase is realized even at finite temperatures. We also determine the spin structure factor, which, in principle, is an experimentally measurable quantity and is the basic tool to map the spectrum of elementary excitations of the system.    
\end{abstract}


\maketitle

\section{Introduction}

The study of Mott insulating states of spin-1/2 systems has a long history in condensed matter physics. The main motivation of these studies lies in their relation to high-$T_c$ superconductivity \cite{lee06a}. Solid state systems are seldom clean, usually lattice defects, vibrations, or simply the complicated nature of the compound obscure the main effects that theoretical models try to grasp. Ultracold atoms on optical lattices seems to provide a nice alternative to study Mott insulators and quantum magnetism, since the optical lattice is free of any defects and there is a great liberty in choosing or even tuning the geometry of the lattice even in situ. In the last decade Mott states of bosonic \cite{greiner02a} as well as fermionic \cite{jordens08a} atoms has been reached experimentally, however, it still remains a challenge to reach quantum magnetism and detect magnetic correlations. The main reason behind the difficulties is the limitation on cooling, since with present experimental methods it is hard to access the temperatures at the scale of the magnetic superexchange, i.e. in the range of few nano Kelvins (for recent progress see \cite{greif13a,duarte13a}). Nevertheless, a series of nice experiments were developed to catch quantum \cite{jordens10a,simon11a,trotzky10a} or even classical \cite{struck11a} magnetism. One possibility is the realization of effective models that can be easier to cool down to sufficiently low temperature. In Ref. \cite{simon11a} tilted Hubbard model was used to mimic an Ising chain, where the empty and double occupied sites represent the spin up and down states, respectively. Another possibility opens by tuning the tunneling amplitude between the neighboring lattice sites to create a staggered-dimerized lattice or a quasi-one-dimensional lattice \cite{trotzky10a}. In these cases the energy scale of the stronger bonds are tuned above the temperature of the cloud, while along the weak bonds the energy is well bellow of it. With this arrangement the system preserves some features of the lower temperature state and can show weak magnetic order. 

High spin Mott insulating states also have been realized experimentally with ytterbium isotopes \cite{taie12a} (see also \cite{taie10a, desalvo10a}).  Ytterbium just as alkaline-earth atoms have 2 electrons on the outer $s$-shell, therefore their total electronic angular momentum is zero, and the total hyperfine spin of the atom comes only from the nuclear spin. These atoms interact essentially via $s$-wave scattering independent of the nuclear spin. As a consequence of the spin independent interaction, alkaline-earth metal systems --- or atoms with equivalent electron shell structure --- can be described with good accuracy even by extremely high SU(N) symmetric models, where N$=2S+1$ is given by the nuclear spin $S$ of the atom. Comparing with the usual two-component electron systems, high spin fermionic systems can show novel magnetic behavior \cite{marston89a,wu10a,wang13a,gorshkov10a}. In the strong atom-atom interaction limit they can provide different multipole orders \cite{tu06a}, valence bond solid (VBS) states, spin liquid (SL) states \cite{wu06a, szirmai11a, hermele09a, hermele11a, szirmai11b,lang13a}, or even chiral spin liquid (CSL) states with non-trivial topology \cite{hermele09a, szirmai11b, hermele11a}. The SU(4) symmetric spin-3/2 system as the simplest case after the usual spin-1/2 electrons, has been studied intensively in the last few years \cite{wu06a, szirmai11a, corboz11a, corboz12a, xu08a, hung11a, cai12a}, mostly on square lattice. On mean-field level the ground state is a VBS state with disconnected resonating valence bond plaquettes, but different numerical results for small systems raises the possibility of a bond-antiferromagnetic columnar dimer state \cite{corboz11a}. On a honeycomb lattice it was found that the pure SU(4) Heisenberg system realizes a spin-orbital liquid phase \cite{corboz12a}, while the addition of next nearest neighbor exchange induces collapse to a tetramerized VBS like state \cite{lajko13a}.

The physics of the SU(6) symmetric spin-5/2 system seems to be even more interesting, as its mean-field ground state is a CSL on a square \cite{hermele09a, hermele11a} and also on a hexagonal lattice \cite{szirmai11b}. The CSL state violates time reversal symmetry and is topologically nontrivial and supports chiral edge states on the boundary of the lattice. The low-energy dynamics is described by 6-flavor spinons (as matter fields) interacting with a U(1) gauge field emerging from the antiferromagnetic correlations. The arising U(1) gauge theory is a dynamical Chern-Simons field theory where the gauge field dynamics is generated by the short distance physics of the underlying fermions \cite{wenbook}. Accordingly, a two-dimensional gas of spin-5/2 alkaline-earth atoms on optical lattices provides a possibility to access experimentally U(1) gauge field dynamics as well as the corresponding non-trivial topological states. This way an alternative quantum simulator of gauge theories can be achieved, which complements other proposals with ultracold atoms \cite{kapit11a,*banerjee13a,*zohar12a,*zohar13a,*zohar13b,*tagliacozzo13a}.

In our previous work the CSL state on a hexagonal lattice has been found to be competing with various VBS states with energies very close (1.5\%) to that of the chiral spin liquid ground state \cite{szirmai11b}. Considering the very small difference between the energies of the low lying states, a stability analysis of these low-energy states is needed to support or even deny their existence.  According to the fact, that experiments are always done at finite temperature, it is also important to see the behavior of the zero temperature mean-field solutions at finite temperatures in order to determine the critical temperatures of the different SL and VBS states, and to see whether the lowest free-energy state changes to another one as the temperature is increased. Finite temperature calculation was made for a similar system in Ref. \cite{hazzard12a} in the high temperature regime, and they managed to describe the metal-insulator transition, but still left open the finite temperature physics far in the Mott insulator regime, where quantum magnetism characterizes the system.

In this paper we address these problems by considering spin-5/2 alkaline-earth atoms loaded into a two-dimensional hexagonal lattice. Supposing a strong on-site repulsive interaction and a filling of one particle per site, the system is described by an effective nearest-neighbor model of SU(6) spin exchange. Hermele, Gurarie and Rey argued in Ref. \cite{hermele09a} that a classical N\'eel-like ground state is unfavorable for SU(N) systems for $\text{N}\ge3$. Therefore we assume that the state is not classically ordered but rather one with a conserved SU(6) symmetry. We analyze the stability and the finite temperature properties of the three lowest lying states within the path integral formalism, which is a reformulation of the Gutzwiller projected variational mean-field theory. We show that the chiral SL state and one of the higher energy VBS states are stable against the displacement from the saddle point, while the staggered-like VBS state occurs to be unstable. The finite temperature analysis shows that by increasing the temperature all three states vanish at the same critical temperature giving way to the paramagnetic phase. The free energy functions never cross each other, the chiral SL state characterizes the system even at finite temperatures, however, in the vicinity of the critical temperature the free energies practically coincide. From experimental aspects it is also important to provide information about the measurable quantities. Accordingly we determine the magnetic structure factor of the three lowest lying states and argue that these states can be distinguished by their fingerprints on the structure factor. We also determine the spectral function of the spin correlator and show how the spectrum of elementary excitations differ for the discussed phases.

The paper is organized as follows. In Sec. \ref{sec:pathint} the model and the path integral formulation of the problem is introduced. With the help of a Hubbard-Stratonovich transformation we introduce slowly varying bosonic fields characterizing the VBS and SL phases. We integrate out the fermion degrees of freedom and arrive to a non-polynomial effective action of the Hubbard-Stratonovich fields. In Sec. \ref{sec:meanfield} we apply the saddle-point expansion which to leading order gives the mean-field solutions of the problem. In Sec. \ref{sec:stability} the next-to-leading order correction is calculated for the effective action in order to study the stability of the saddle-point solutions. In Sec. \ref{sec:structfact} the spin-spin correlation function and its spectral function are calculated and compared for the different saddle-point solutions. Sec. \ref{sec:summary} is the summary. Some calculations are moved to the Appendix.

\section{Path integral formulation of SU(N) magnetism}
\label{sec:pathint}

We consider a system of ultracold spin-5/2 $^{173}\mathrm{Yb}$ atoms on an optical lattice with hexagonal structure. For a single Yb atom the angular momentum of the electrons is zero since the outer shell is an $s$-shell with two opposite spin electrons. Therefore the total spin of an Yb atom is given by its nuclear spin alone and as a consequence atomic collisions become spin independent. At low temperatures and when the optical lattice is sufficiently deep the atoms occupy the lowest bands of the individual sites and the system can be described by a Hubbard model generalized to 6 components.
\begin{equation}
 H_{\text{Hub}} = -t \sum_{\langle i,j\rangle,\alpha} \left( c_{i \alpha}^{\dagger} c_{j \alpha} + \text{H.c.} \right)+
\frac{U}{2} \sum_i n_i (n_i - 1),
 \label{Hubbard_psinko}
\end{equation}
where the first sum is over nearest neighbor pairs while the second sum runs through the sites of the whole lattice. The operator $c^{\dagger}_{i \alpha}$ creates a fermion at site $i$ with spin component $\alpha\in\lbrace-5/2\ldots5/2\rbrace$, and $t$ represents the hopping amplitude. In the second term $n_i = \sum_\alpha c_{i \alpha}^{\dagger} c_{i \alpha}$ denotes the occupation number of site $i$, and $U>0$ is the strength of the repulsive, SU(6) symmetric on-site interaction.      

We consider the case of 1/6 filling, where the number of lattice sites is equal to the number of atoms. We also assume strong on-site interaction, $U\gg t$. This limit can be already achieved in standard tight optical lattices, as recently demonstrated by the LENS group \cite{fallani}, where the Tonks parameter $\gamma\approx 5$ in the 1-dimensional ytterbium gas was reached, indicating a strongly correlated regime. Another idea is to use supertight lattices employing nanotechnology, as proposed in Ref. \cite{gullans12a}. This limit can in principle also be realized in experiments with Feshbach or optical Feshbach resonances \cite{taie12a}, although this method is experimentally very challenging.

In the exact limit of $t/U=0$, the hopping term vanishes from Eq. \eqref{Hubbard_psinko} and the lattice gas becomes completely static, with a highly degenerate ground state, which is a Mott insulator state with exactly one particle at every site. At low temperatures and for $U/t\gg 1$, but finite, one can use perturbation theory in $t/U$ up to second order to resolve the degeneracy and to arrive to an effective spin exchange Hamiltonian (similarly as in the case of the standard Fermi-Hubbard model \cite{auerbachbook})
\begin{equation}
H = - J \sum_{\langle i,j\rangle,\alpha,\beta} c^{\dagger}_{i \alpha} c_{j \alpha} c^{\dagger }_{j \beta} c_{i\beta} \, ,
\label{Heisenberg_sinko}
\end{equation}
where $J=4t^2/U$ is an antiferromagnetic coupling. This effective Hamiltonian describes processes conserving the unit population at every site and only allows the exchange of spins between nearest neighbor sites. One can considered this model as a generalization of the SU(2) Heisenberg model to SU(6) spins, since the Hamiltonian \eqref{Heisenberg_sinko} can also be expressed as $H=J\sum_{\langle i,j\rangle,\alpha\beta}S_{i,\alpha\beta}S_{j,\beta\alpha}$, where the ``artificial'' SU(6) spin operators are given by
\begin{equation}
\label{eq:su_n_spin}
S_{i,\alpha\beta}=c_{i,\alpha}^\dagger c_{i,\beta}.
\end{equation}
These SU(6) generators are not to be confused with the 6 dimensional representation of the original atomic spin operators. The $\mathrm{6}^2=36$ spin operators at every site, given by Eq. \eqref{eq:su_n_spin}, span the SU(6) Lie algebra, i.e. operators on the same lattice site satisfy the commutator
\begin{equation}
\label{eq:su_n_cr}
[S_{i,\alpha\beta},S_{i,\gamma\delta}]=S_{i,\gamma\beta}\,\delta_{\alpha\delta}-S_{i,\alpha\delta}\,\delta_{\gamma\beta},
\end{equation}
while operators on different lattice sites commute with each other. According to the one particle per site constraint, the operator sum $\sum_{\alpha}S_{i,\alpha\alpha}=\sum_{\alpha}c_{i,\alpha}^\dagger c_{i,\alpha}=1$. It means that this sum of the SU(6) spin operators commutes with all other SU(6) operators, therefore the number of independent generators is actually $36-1=35$. Furthermore, it means that the dimension of the local Hilbert space at site $i$ is 6, and thus we work with the fundamental representation of SU(6). One should stress, however,  that in the following we do not make use of the description in terms of local spins, allowing instead for Fermi pairings on the links, similarly as in the case of slave-boson theories \cite{lee06a}.

The Hamiltonian \eqref{Heisenberg_sinko} has local U(1) gauge invariance, namely it is invariant under the following local gauge transformation
\begin{subequations}
\label{fermion_sym_sinkovicz}
	\begin{align}
        c_{i \alpha}& \to c_{i\alpha} e^{i \Theta_i},\\
        c^\dagger_{i \alpha}& \to c^\dagger_{i\alpha} e^{-i \Theta_i},
	\end{align}
\end{subequations}
with the arbitrary position and time dependent real parameters $\Theta_i(t)$.

We discuss finite temperature properties of the system. The key quantity is the canonical partition function at inverse temperature $\beta$ which is evaluated in the imaginary time path-integral formalism \cite{altlandsimons}
\begin{equation}
\label{eq:partfun}
Z(\beta) = \int D[\overline{c},c] e^{-S[\overline{c},c]} \prod_{i,\tau} \delta\left(\sum_\alpha\overline{c}_{i \alpha}c_{i \alpha} - 1\right).
\end{equation}
The delta functions in the integrand assure the one-particle constraint at every site and for every imaginary time
$\tau = -i t$. We use $\hbar = 1$ units, and  $\overline{c}$ ($c$) are Grassmann numbers associated to $c^\dagger$ ($c$). The action is given by
\begin{equation}
\label{eq:action}
S[\overline{c},c] = \int_0^\beta d \tau \left( \sum_{i,\alpha} \overline{c}_{i \alpha} \partial_\tau c_{i \alpha} + H \right).
\end{equation}
Since the Hamiltonian \eqref{Heisenberg_sinko} is quartic in the fermion fields we need to rely on an approximation scheme. According to the pioneering works by Marston and Affleck on the general SU(N) Hubbard models \cite{marston89a} and the more recent analysis of Hermele, Gurarie and Rey on SU(N) symmetric models realized with ultracold atoms \cite{hermele09a} we consider only spin liquid states, i.e. where the global SU(6) symmetry is not broken. In this case, with the help of a Hubbard-Stratonovich (HS) transformation \cite{altlandsimons}, new, slowly varying fields are introduced.
\begin{multline}
\label{eq:HS}
\exp\bigg(J\int_0^\beta d\tau\sum_{\langle i,j\rangle,\alpha,\beta}\overline{c}_{i,\alpha}c_{j,\alpha}\overline{c}_{j,\beta}c_{i,\beta}\bigg)\\
=\int D[\chi^*,\chi]\exp\bigg\lbrace
-\int_0^\beta d\tau\sum_{\langle i,j\rangle}\Big[\frac{1}{J}|\chi_{ij}|^2\\
-\sum_\alpha\big(\chi_{ij}\overline{c}_{j,\alpha}c_{i,\alpha}
-\chi_{ij}^*\overline{c}_{i,\alpha}c_{j,\alpha}\big)
\Big]
\bigg\rbrace
\end{multline}
The Hubbard-Stratonovich field, $\chi_{ij}$, lives on the links between adjacent sites and are complex, furthermore $\chi_{ij}^*=\chi_{ji}$. Finally, another bosonic field, $\varphi_i$, is introduced in order to cast the delta functions also to a Gaussian form by
\begin{multline}
\label{eq:deltarep}
\prod_{i,\tau} \delta\left(\sum_\alpha\overline{c}_{i \alpha}c_{i \alpha} - 1\right)\\
=\int D[\varphi]\exp
\bigg[
-\int_0^\beta d\tau\sum_i\varphi_i\bigg(\sum_\alpha\overline{c}_{i,\alpha}c_{i,\alpha}-1\bigg)
\bigg].
\end{multline}
The bosonic field $\varphi_i$ is purely imaginary for the proper Fourier representation of the delta functions. However, we conceal its imaginary nature, because it is very suggestive in the final form of the action, as playing the role of a scalar potential for the fermions. Combining Eqs. \eqref{Heisenberg_sinko}, \eqref{eq:partfun}, \eqref{eq:action}, \eqref{eq:HS} and \eqref{eq:deltarep} the partition function takes the form  
\begin{equation}
\label{eq:partfunctot}
Z(\beta) = \int D [\overline{c},c,\chi^*,\chi,\varphi] e^{- S_{\text{tot}}[\overline{c},c,\chi^*,\chi,\varphi]} \, ,
\end{equation}
with the total action
\begin{multline}
S_{\text{tot}}[\overline{c},c,\chi^*,\chi,\varphi] = \int_0^\beta \textrm{d}\tau \Bigg\lbrace \sum_{i,\alpha} \overline{c}_{i \alpha} ( \partial_\tau + \varphi_i ) c_{i \alpha} \\ 
-\sum_{\langle i,j\rangle} \bigg[\sum_\alpha\bigg( \chi_{ij} \overline{c}_{j\alpha} c_{i \alpha} + \mathrm{H.c.}\bigg) - \frac{1}{J} |\chi_{ij}|^2 \bigg]- \sum_i \varphi_i \Bigg\rbrace.
\label{total_action_sinkovicz}
\end{multline}

The total action \eqref{total_action_sinkovicz} is also U(1) gauge invariant with the additional transformation rules:
\begin{subequations}
\label{eqs:gauge_freedom}
\begin{align}
        \chi_{ij}& \to \chi_{ij} e^{i(\Theta_j - \Theta_i)},\label{eq:gauge_chi}\\
        \varphi_i & \to \varphi_i - i \partial_\tau \Theta_i.\label{eq:gauge_phi}
\end{align}
\end{subequations}
Thus the phase of $\chi$ transforms as a vector potential, and $\varphi$ transforms as a scalar potential. With the introduction of these fields the fermion action \eqref{total_action_sinkovicz} is quadratic, and one can think about the system as noninteracting fermions hopping on the sites of the hexagonal lattice immersed into a scalar potential $\varphi_i$ and vector potential $\arg \chi_{ij}$. One has to keep in mind that the appearing gauge fields are not static: there is functional integration over all the possible $\chi$ and $\varphi$ configurations. Indeed, the integration over the gauge degrees of freedom is the essence of the path-integral formulation of the problem as it renders the expectation values of all non gauge invariant quantities to zero. This way the mean value of those operators which violate the one particle per site constraint are annulled \cite{baskaran88a,arovas88a}. 

\subsection{Choice of the unit cell: six-sublattice ansatz}

\begin{figure}[tb]
\begin{center}
  \includegraphics{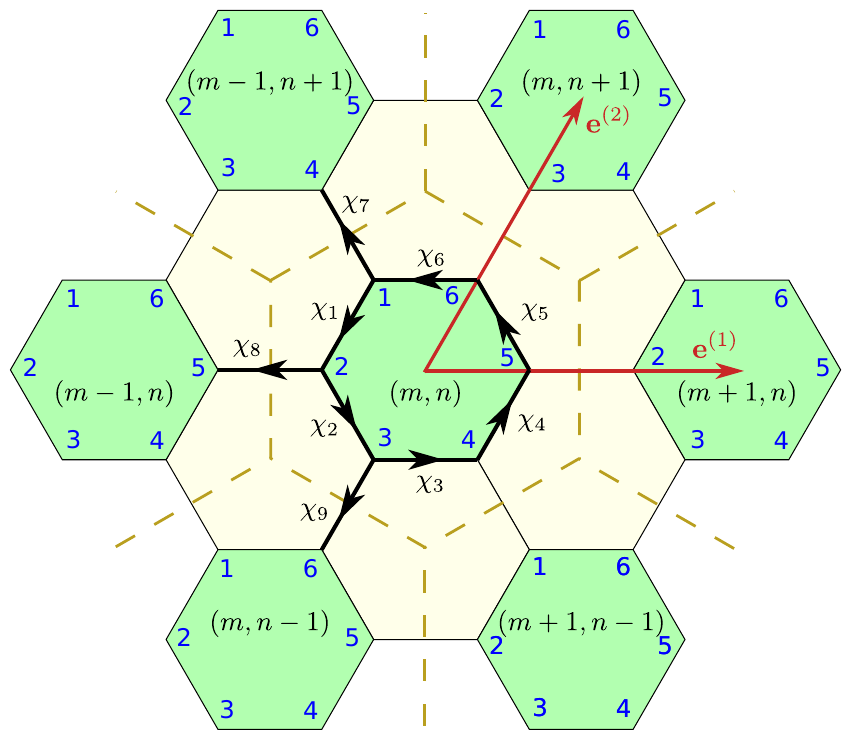}
  \caption{(Color online) Illustration of the unit cell considered here. The unit cell is surrounded by the dashed line and contains 6 sites. The numbers close to the nodes (blue in the color version) label the sublattice index (the sites inside the unit cell), while the $(m,n)$ pair indexes the unit cell itself. There are 6 Hubbard-Stratonovich fields inside each unit cell labeled by $\chi_1\ldots\chi_6$ and 3 other independent ones between the neighboring unit cells $\chi_7\ldots\chi_9$. We also show the elementary lattice vectors connecting the neighboring unit cells.}
  \label{fig:unitcell}
\end{center}
\end{figure}

In order to use momentum space representation on the honeycomb lattice one needs to introduce a nontrivial unit cell. The simplest choice is to use two sublattices, which is the case used in graphene, or generally in systems with two component fermions on hexagonal lattices. In our six component fermion case, especially when we restrict ourselves to $1/6$ filling, a more general, six sublattice ansatz is more favorable. Our choice of the unit cell is depicted in Fig. \ref{fig:unitcell}. A site $\mathbf{r}_{m,n,s}$ is indexed by three integers, from which $m$ and $n$ are selecting the unit cell (with a coordinate $\mathbf{r}_{m,n}=m\,\mathbf{e}^{(1)}+n\,\mathbf{e}^{(2)}$ pointing to the center of the cell), and another integer $s\in\lbrace1\ldots6\rbrace$ selecting the sublattice inside the cell. The elementary lattice vectors $\mathbf{e}^{(1)}$ and $\mathbf{e}^{(2)}$ point from the center of a unit cell to the centers of the two neighboring unit cells as illustrated in Fig. \ref{fig:unitcell}. The fermion fields $c_\alpha(\mathbf{r}_{m,n,s})$ are arranged into the 6 component vector $c_{s,\alpha}(\mathbf{r}_{mn})=[c_\alpha(\mathbf{r}_{mn1}), c_\alpha(\mathbf{r}_{mn2}), \ldots c_\alpha(\mathbf{r}_{mn6})]_s$.

We use plane wave basis for one particle states, therefore the fermionic field operator is expressed as:
\begin{equation}
    c_{s,\alpha}(\mathbf{r}_{m,n},\tau) = \frac{1}{\sqrt{V\beta}} \sum_{\mathbf{k},l} c_{s,\alpha}(\mathbf{k},i\omega_l) e^{i(k_1m+k_2 n - \omega_l \tau)},
\end{equation}
where $\omega_l=(2l+1)\pi/\beta$ is the Matsubara frequency for fermions, and $\mathbf{k}=k_1\mathbf{f}^{(1)}+k_2\mathbf{f}^{(2)}$ is the wavenumber in reciprocal space spanned by $\mathbf{f}^{(1)}$ and $\mathbf{f}^{(2)}$. We use the normalization $(\mathbf{f}^{(i)},\mathbf{e}^{(j)})=\delta_{ij}$.

With our choice of the unit cell we have 6 HS fields inside the cell $(m,n)$ (see Fig. \ref{fig:unitcell}.), $\chi_1(\mathbf{r}_{mn})\ldots\chi_6(\mathbf{r}_{mn})$ and another 3 connecting the unit cell $(m,n)$ to its 3 neighbors to the left, $\chi_7\ldots\chi_9$. The HS fields are bosonic, therefore in their Matsubara representation,
\begin{equation}
\label{eq:FourerHS}
    \chi_i(\mathbf{r}_{m,n},\tau) = \frac{1}{V\beta} \sum_{\mathbf{k},l}               
    \chi_i(\mathbf{k},i\nu_l) e^{i(k_1m+k_2 n - \nu_l \tau)},
\end{equation}
the frequencies are $\nu_l = 2l\pi/\beta$. The $\varphi_s(\mathbf{r}_{mn})$ fields are over the sites, therefore we have 6 of them for each unit cell. A similar transformation to Eq. \eqref{eq:FourerHS} is understood for them, with $\chi_i$ replaced by $\varphi_s$. Here and from now on $i=1\ldots9$ indexes the 9 different $\chi_i$ fields and $s=1\ldots6$ denotes the sublattice and therefore indexes the $\varphi_s$ fields.

The total action Eq. \eqref{total_action_sinkovicz} in momentum space is evaluated by some straightforward algebra to
\begin{multline}
S_\text{tot}[\overline{c},c,\chi^*,\chi,\varphi]
= -\sum_s \varphi_s(\hat{q}=0)\\
+\frac{1}{JV\beta} \sum_{\hat{q},i} |\chi_i(\hat{q})|^2
-\sum_{\hat{k},\hat{q},s,s',\alpha}\!\overline{c}_{s'\alpha}(\hat{k}+\hat{q})  G_{s's}^{-1}(\hat{k},\hat{q}) c_{s \alpha}(\hat{k}),
\label{action_matrsu_plan_sinko}
\end{multline}
where we have used the shorthand notations $\hat{k}\equiv(\mathbf{k},i\omega_n)$, and $\hat{q}\equiv(\mathbf{q},i\nu_m)$. In the sum $\alpha$ is over the spin components of the fermions, while $i=1\ldots9$, and $s,s'=1\ldots6$. $G^{-1}_{s',s}(\hat{k},\hat{q})$ is the inverse of the fermion propagator depending on the boson fields. Its explicit form is given by 
\begin{equation}
G^{-1}_{s's}(\hat{k},\hat{q}) = i \omega_n \delta_{\mathbf{q},\mathbf{0}}\delta_{m,0} \delta_{s',s} - H_{s's}(\hat{k},\hat{q}),
\label{inv_prop_sinko}
\end{equation}
and 
\begin{widetext}
\begin{equation}
\label{eq:Hmatrix}
H(\hat{k},\hat{q})\!=\!\frac{-1}{\beta V}
\!\left[\!
\begin{array}{c c c c c c}
-\varphi_1(\hat{q})&\chi_1(\hat{q})&0&\chi_7(\hat{q})e^{-i(k_1-k_2)}&0&\chi_6^*(-\hat{q})\\
\chi_1^*(-\hat{q})&-\varphi_2(\hat{q})&\chi_2(\hat{q})&0&\chi_8(\hat{q})e^{-ik_1}&0\\
0&\chi_2^*(-\hat{q})&-\varphi_3(\hat{q})&\chi_3(\hat{q})&0&\chi_9(\hat{q})e^{-ik_2}\\
\chi_7^*(-\hat{q})e^{i(k_1+q_1-k_2-q_2)}&0&\chi_3^*(-\hat{q})&-\varphi_4(\hat{q})&\chi_4(\hat{q})&0\\
0&\chi_8^*(-\hat{q})e^{i(k_1+q_1)}&0&\chi_4^*(-\hat{q})&-\varphi_5(\hat{q})&\chi_5(\hat{q})\\
\chi_6(\hat{q})&0&\chi_9^*(-\hat{q})e^{i(k_2+q_2)}&0&\chi_5^*(-\hat{q})&-\varphi_6(\hat{q})
\end{array}\!\right].
\end{equation}
\end{widetext}
All the quantities $\varphi_s$ and $\chi_i$ depend on the transferred momentum $\mathbf{q}$ and Matsubara frequency $\nu_m$. In the exponentials $k_1, k_2, q_1$ and $q_2$ denote the respective components of the wavenumbers. Note that $H(\hat k,\hat q)$  depends only the space-like component of $\hat k$, but depends both on $\mathbf{q}$ and $\nu_m$ through the bosonic fields.

The total action Eq. \eqref{action_matrsu_plan_sinko} is quadratic in the fermion fields, therefore in Eq. \eqref{eq:partfunctot} the functional integral over them can be readily evaluated to
\begin{equation}
\label{eq:Zboson}
Z(\beta)=\int D[\chi^*,\chi,\varphi]\,e^{-S_\text{eff}[\chi^*,\chi,\varphi]},
\end{equation}
with an effective action
\begin{multline}
\label{eq:Seff}
S_\text{eff}[\chi^*,\chi,\varphi] = - \sum_s \varphi_s(\hat{q}=0) + \frac{1}{JV\beta} \sum_{\hat{k},i} |\chi_i(\hat{k})|^2\\
- 6\,\mathrm{tr}[\ln (\beta G^{-1})]
\end{multline}
where the trace is a sum over $\hat{k}$ and $s$. The factor 6 comes from the summation over the spin index $\alpha$.

Equations \eqref{eq:Zboson} and \eqref{eq:Seff} are the main results of the general path integral formulation of SU(N) magnetism for non-classically ordered antiferromagnetic states \cite{marston89a,hermele11a}. The effective action Eq. \eqref{eq:Seff} is a functional of the slowly varying (both in space and in imaginary time) bosonic fields $\chi$ and $\varphi$. In Eq. \eqref{eq:Zboson} $e^{-S_{\text{eff}}}$ serves as the weight function of the possible field configurations. If this weight function was known, all the necessary correlation functions of the fields and also of the original fermion operators could be calculated exactly.  

\section{Saddle-point approximation}
\label{sec:meanfield}

\subsection{Derivation of the saddle-point equations}

The effective action \eqref{eq:Seff} for the boson fields is non-polynomial. In order to evaluate it one needs to rely on an approximation scheme, the saddle-point evaluation method: we assume that the probability distribution, $e^{-S_{\text{eff}}}$, in the field integral \eqref{eq:Zboson} is dominated by its maximum values, and in the vicinity it is close to a Gaussian. We proceed by expanding the fields $\chi$ and $\varphi$ around the homogeneous saddle-point configurations and by considering the fluctuations around them,
\begin{subequations}
\label{eqs:SPexpansion}
\begin{align}
\chi_i(\hat{q}) &=\beta V \bar\chi_i \delta_{\hat{q},0}+\delta\chi_i(\hat{q}),\\
\varphi_s(\hat{q})&=\beta V \bar\varphi_s \delta_{\hat{q},0}+\delta\varphi_s(\hat{q}).
\end{align}
\end{subequations}
Here and in the following $\delta_{\hat{q},0}=\delta_{\mathbf{q},0}\delta_{m,0}$. The complex numbers $\bar\chi$  and $ \bar\varphi $ are the homogeneous stationary points: the functional derivative of $S_\text{eff}$ with respect to $\delta\chi$ and $\delta\varphi$ has to vanish.
\begin{equation}
\left.\frac{\delta S_\text{eff} [\delta\chi^\ast,\delta\chi,\delta\varphi]}{\delta\phi}\right|_{\delta\chi=\delta\chi^*=\delta\varphi=0}=0
\label{saddle_point_alex}
\end{equation}
where we have introduced a 24 component vector $\phi_\mu$ composed of the fluctuations. The first 18 elements are the complex fields $\delta\chi$ and their complex conjugates, and the last 6 elements are the $\delta\varphi$ fields: $\phi_\mu(\hat{q})=[\delta\chi_i(\hat{q}), \delta\chi_i^*(-\hat{q})),\delta\varphi_s(\hat{q})]_\mu$.

With the decomposition \eqref{eqs:SPexpansion} of the fields $\chi$ and $\varphi$, the inverse of the fermionic Green's function \eqref{inv_prop_sinko} is split to two parts 
\begin{equation}
G^{-1}_{s's}(\hat{k},\hat{q}) = G^{-1}_{(0)s's}(\hat{k},\hat{q})-\Sigma_{s's}(\hat{k},\hat{q}),
\label{inv_prop_mean_value_alex}
\end{equation}
where the inverse of the saddle-point Green's function, $G_{(0)}^{-1}$, contains only the saddle-point values of the HS fields, while the self-energy, $\Sigma$ contains the fluctuations of the HS fields (their explicit forms are given below). 
With such a separation a systematic expansion around the saddle point can be done in terms of powers of $\Sigma$.
The effective action \eqref{eq:Seff} can be expanded in powers of the fluctuations around the mean field:
\begin{subequations}
\label{eqs:Seffexp}
\begin{align}
S_{\text{eff}}&=S_0+S_1+S_2+6\sum_{n=1}^{\infty}\frac{\mathrm{tr}\;(G^0\Sigma)^n}{n},\label{eq:Seffexp}\\
S_0&=-\beta V\sum_{s}\bar\varphi_s+\frac{\beta V}{J}\sum_i | \bar\chi_i |^2 -6\,\mathrm{tr}\log(\beta G^{-1}_0),  \label{eq:Seff0}\\
S_1&=-\sum_s\delta\varphi_s(\hat{q}=0)+\frac{1}{J}\sum_i \left[  \bar\chi^{*}_i\delta\chi_i(\hat{q}=0) + \text{H.c.}  \right],\label{eq:Seff1}\\
S_2&=  \frac{1}{J\beta V}\sum_{i,\hat{q}}\,\delta\chi^{*}_i(\hat{q})\delta\chi_i(\hat{q}),\label{eq:Seff2}
\end{align}
\end{subequations}
where $S_0,S_1,S_2$ contains the fluctuations of the fields with zeroth, first, and second order, respectively, and the logarithm in Eq. \eqref{eq:Seff} is expanded in powers of the self-energy $\Sigma$ as:
\begin{multline}
\mathrm{tr}\log(\beta G^{-1})=\mathrm{tr}\log\left[  \beta(G_{(0)}^{-1}-\Sigma)  \right]\\
=\mathrm{tr}\log\left(\beta G^{-1}_{(0)}\right)+\sum_{n=1}^{\infty}\frac{\mathrm{tr}(G_{(0)}\Sigma)^n}{n}.
\end{multline}
The evaluation of $\mathrm{tr}(G_{(0)}\Sigma)^n$ via a systematic Feynman diagram method is detailed in Appendix \ref{appendix}. 

Now the saddle-point equations \eqref{saddle_point_alex} are cast to a more direct form by collecting the first order contributions to the effective action, and therefore
\begin{equation}
\label{eq:SPeqsformal}
\frac{\partial S_1}{\partial \phi_\mu(\hat{q})}  +  6\,\mathrm{tr} \left(  G_{(0)} \frac{\partial \Sigma}{\partial \phi_\mu(\hat{q})}  \right)=  0.
\end{equation}
These equations provide the self consistent equations for the mean-field (i.e. saddle-point) solutions. In order to solve them we need the explicit form of the saddle-point Green's function $G_{(0)}$, and the self-energy $\Sigma$.

The saddle-point Green's function is diagonal in momentum space, and in the Matsubara frequencies:
\begin{equation}
G^{-1}_{(0)s's}(\hat{k},\hat{q}) = G^{-1}_{(0)s's}(\hat{k})\delta_{\hat{q},0},
\end{equation}
with
\begin{equation}
\label{eq:freegreensfunc}
G^{-1}_{(0)s's}(\hat{k})\equiv G^{-1}_{(0)s's}(\mathbf{k},i\omega_n)=\left( i \omega_n\delta_{s',s} - H_{s's}^{(0)}(\mathbf{k})\right).
\end{equation}
The mean-field fermion Hamiltonian $H_{s's}^{(0)}(\mathbf{k})\delta_{\hat{q},0}$ is obtained by replacing $\chi_i(\hat{q})$ and $\varphi_s(\hat{q})$ in Eq. \eqref{eq:Hmatrix} with $\beta V \bar\chi_i \delta_{\hat{q},0}$ and $\beta V \bar\varphi_s \delta_{\hat{q},0}$. The $H_{s's}^{(0)}(\mathbf{k})$ Hamiltonian can be easily diagonalized for all $\mathbf{k}$ momentum
\begin{equation}
\label{eq:H0diag}
\sum_s H_{s's}^{(0)}(\mathbf{k}) v_s^{(a)}(\mathbf{k})=\varepsilon^{(a)}_{\mathbf{k}} v_s^{(a)}(\mathbf{k}),
\end{equation}
with eigenvalues $\varepsilon^{(a)}_{\mathbf{k}}$ and eigenvectors $v_s^{(a)}(\mathbf{k})$, where the eigenvalue index is $a\in\lbrace1\ldots6\rbrace$. With the help of Eqs. \eqref{eq:freegreensfunc} and \eqref{eq:H0diag} the saddle-point Green's function is expressed as
\begin{equation}
\label{eq:freeprop}
G_{(0)s's}(\hat{k})\equiv G_{(0)s's}(\mathbf{k},i\omega_n)=\sum_a\frac{v_{s'}^{(a)}(\mathbf{k})\,v_s^{(a)*}(\mathbf{k})}{i\omega_n-\varepsilon^{(a)}_{\mathbf{k}}}.
\end{equation}
Note, that the saddle-point Green's function depends on the saddle-point values of the fields, $\bar \chi$ and $\bar \varphi$, through the eigenvalues and eigenvectors of the matrix $H^{(0)}$.

The self-energy is obtained by replacing $\chi_i(\hat{q})$ and $\varphi_s(\hat{q})$ in Eq. \eqref{eq:Hmatrix} with their fluctuations $\delta\chi_i (\hat{q})$ and $\delta\varphi_s (\hat{q})$. It can be compactly written as
\begin{multline}
\label{eq:sigma}
\Sigma_{s's}(\hat{k},\hat{q})=\frac{-1}{\beta V}\sum_{i=1}^9\Big[\delta_{s',\beta_i}\delta_{s,\alpha_i}e^{-i\gamma_i(\mathbf{k})}\delta\chi_i(\hat{q})\\
+\delta_{s',\alpha_i}\delta_{s,\beta_i}e^{i\gamma_i(\mathbf{k}+\mathbf{q})}\delta\chi_i^*(-\hat{q})\Big]
+\frac{1}{\beta V}\sum_{r=1}^6 \delta_{s'r}\delta_{sr}\delta\varphi(\hat{q}).
\end{multline}
The specific form of the newly introduced indices $\alpha_i$ and $\beta_i$ are given in Table \ref{tab:vertices} together with the phase factor $\gamma_i(\mathbf{k})$ describing the momentum dependence of the inter unit cell links (compare with Eq. \eqref{eq:Hmatrix}).
\begin{table}[tb]
\begin{tabular}{c c c c|c c c c|c c c c}
\hline
\hline
$j$&$\alpha_j$&$\beta_j$&$\gamma_j(\mathbf{k})$&$j$&$\alpha_j$&$\beta_j$&$\gamma_j(\mathbf{k})$&$j$&$\alpha_j$&$\beta_j$&$\gamma_j(\mathbf{k})$\\
\hline
1&2&1&0&4&5&4&0&7&4&1&$k_1-k_2$\\
2&3&2&0&5&6&5&0&8&5&2&$k_1$\\
3&4&3&0&6&1&6&0&9&6&3&$k_2$\\
\hline
\hline
\end{tabular}
\caption{The incoming $\alpha_j$ and outgoing $\beta_j$ fermion line sublattice indices for vertex $j$ and the phase factor $\gamma_j(\mathbf{k})$. See Appendix \ref{appendix}.}
\label{tab:vertices}
\end{table} 

Equation \eqref{eq:SPeqsformal} can be cast to an explicit form now. The first term is easily evaluated with the help of Eq. \eqref{eq:Seff1}. In the second term the trace is a sum over the wavenumber $\mathbf{k}$, Matsubara frequency $\omega_n$ and the sublattice index $s$. 
Using the "free" propagator \eqref{eq:freeprop} and performing the sum for the Matsubara frequencies one arrives to
\begin{multline}
\label{eq:firstordertotrace}
\mathrm{tr}\left(G_{(0)}\frac{\partial \Sigma}{\partial \phi_\mu(\hat{q})}\right)=\sum_{\hat{k},s,s'}G_{(0)s,s'}(\hat{k})\frac{\partial \Sigma_{s',s}(\hat{k},\hat{q})}{\partial \phi_\mu(\hat{q})}\\
=\sum_{\mathbf{k},a,s,s'}\beta\frac{v^{(a)}_s(\mathbf{k})\,v^{(a)*}_{s'}(\mathbf{k})}{1+e^{\beta\varepsilon^{(a)}_\mathbf{k}} }\,\frac{\partial \Sigma_{s',s}(\hat{k},\hat{q})}{\partial \phi_\mu(\hat{q})}.
\end{multline}
The derivative of $\Sigma$ can also be readily evaluated with the help of its compact form Eq. \eqref{eq:sigma}. 
After some straightforward algebra one arrives to the explicit form of the self-consistency equations \eqref{eq:SPeqsformal}:
\begin{subequations}
\label{saddle_eq_alex}
\begin{align}
\bar\chi_j&=\frac{6J}{V}\sum_{\mathbf{k},a} e^{i\gamma_j(\textbf{k})} \frac{v^{(a)}_{\beta_j}(\mathbf{k})\,v^{(a)*}_{\alpha_j}(\mathbf{k})}{1+e^{\beta\varepsilon^{(a)}_\mathbf{k}} },&\text{for } j\in\lbrace1\ldots9\rbrace , \label{eq:chieq}\\
1&=\frac{6}{V}\sum_{\mathbf{k},a}\frac{v^{(a)}_s(\mathbf{k})\,v^{(a)*}_s(\mathbf{k})}{1+e^{\beta\varepsilon^{(a)}_\mathbf{k}} },&\text{for } s\in\lbrace1\ldots6\rbrace . \label{eq:occupeq}
\end{align}
\end{subequations}
Only 15 of the 24 equations are independent, the remaining 9 equations are complex conjugates of Eqs. \eqref{eq:chieq}. 
The solutions of Eqs. \eqref{saddle_eq_alex} provide the mean-field solutions of the problem. At finite temperature each solution is characterized by its free energy, and the configuration with the lowest free energy dominates the partition function. 
The free energy, $F=-k_B T \log Z(\beta)$, at mean-field level, i.e. by neglecting quantum fluctuations, is given by
\begin{multline}
F_{\text{mf}}(T,V)=\frac{S_0}{\beta}=-V\sum_s\bar\varphi+\frac{V}{J}\sum_i | \bar\chi |^2\\
-\frac{6}{\beta}\sum_{\textbf{k},a}\mathrm{log}\left(1+e^{-\beta\varepsilon^{(a)}_{\textbf{k}}}\right),
\end{multline}
where we have performed the sum for the Matsubara frequencies \cite{altlandsimons}
\begin{multline}
\mathrm{tr}\log(\beta G^{-1}_0)=\sum_{\mathbf{k},n,a} \log\left(i\beta\omega_n- \beta\epsilon^{(a)}_{\mathbf{k}}\right)\\
=\sum_{\mathbf{k},a} \log\left( 1+ e^{-\beta\epsilon^{(a)}_{\mathbf{k}}} \right).
\end{multline}
In order to describe the possible competition of various configurations we determine the three lowest lying solutions. Their basic properties are discussed in the next subsection.

\begin{figure*}
\begin{center}
  \includegraphics{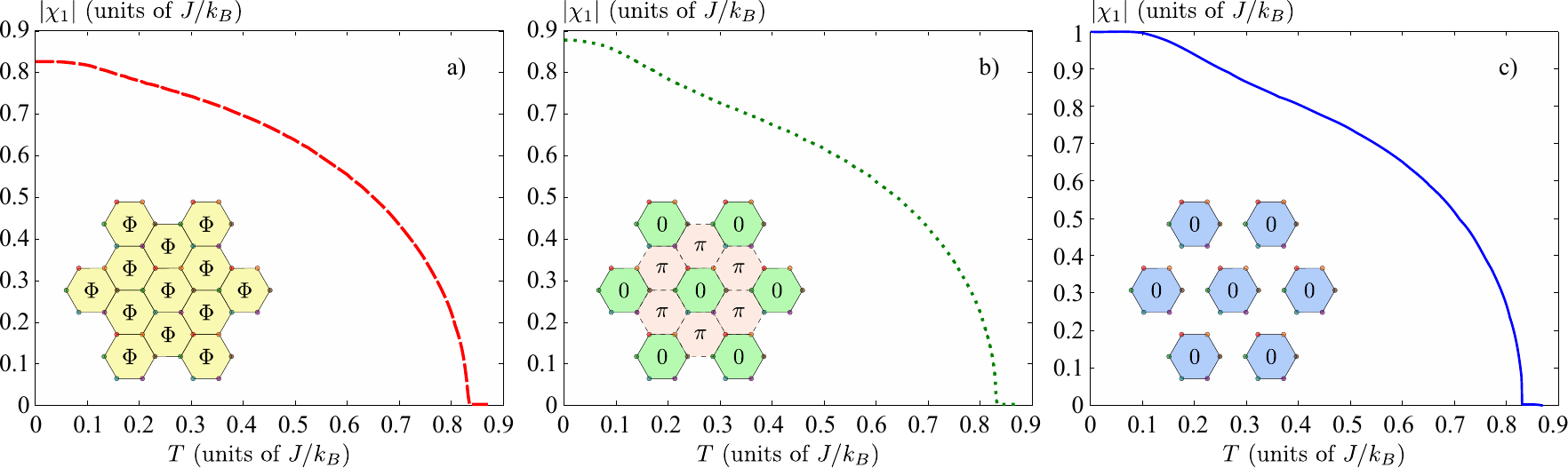}
  \caption{(Color online) The saddle-point solutions for the three lowest free energy spin liquid states. a) The chiral spin liquid state. b) The quasi plaquette state. c) The plaquette state.}
  \label{fig:mfsols}
\end{center}
\end{figure*}

It is worth to emphasize that we have used the term ``mean field'' throughout this section as a synonym for ``saddle point''. Eventually, the mean-field approximation in Ref. \cite{szirmai11b} is equivalent to the calculation performed here in the limit when $\beta=\infty$. In the mean-field calculation the $\chi_{ij}$ mean fields are the fermion correlators $\chi_{ij}=\sum_\alpha\langle c_{i\alpha}^\dagger c_{j\alpha}\rangle$. Here they are HS fields and their saddle-point values correspond to the mean fields. The advantage of the path-integral method beyond its transparency is its straightforward application for finite temperatures just as when going beyond the saddle-point approximation as we will see in Sec. \ref{sec:stability}.

\subsection{Finite temperature mean-field solutions}

In this section we present the lowest free energy solutions of Eqs. \eqref{saddle_eq_alex}. First we have to note, that there are infinitely many solutions of the coupled equations due to the gauge freedom Eqs. \eqref{eqs:gauge_freedom}; from any set of saddle-point solutions we can generate new ones with the same free energy by applying an arbitrary $\tau$ independent transformation according to Eq. \eqref{eq:gauge_chi}. In order to factor out this trivial gauge freedom, we distinguish the saddle-point solutions by the elementary Wilson loops, i.e. those solutions are considered to be equivalent whose elementary Wilson loops are equal. An elementary Wilson loop is a product of the $\chi$ fields around an elementary plaquette of the lattice. 
The unit cell consists three elementary plaquettes, and in the saddle-point approximation the unit cell is repeated periodically. Therefore there are three independent elementary plaquettes only. One of them is the plaquette corresponding to the central hexagon of the unit cell, its Wilson loop is $\Pi_1=\bar\chi_1\,\bar\chi_2\,\bar\chi_3\,\bar\chi_4\,\bar\chi_5\,\bar\chi_6$ (see Fig. \ref{fig:unitcell}).  The other two can be chosen from the neighboring plaquettes with the requirement that they have to be independent. One of them can be the one which is between the unit cells $(m,n)$, $(m-1,n)$ and $(m-1,n+1)$, its Wilson loop is $\Pi_2=\bar\chi_7\,\bar\chi_3^*\,\bar\chi_4^*\,\bar\chi_5^*\,\bar\chi_8^*\,\bar\chi_1^*$, and the third one can be the one between $(m,n)$ and $(m-1,n)$ and $(m,n-1)$, with a Wilson loop $\Pi_3=\bar\chi_8\,\bar\chi_4^*\,\bar\chi_5^*\,\bar\chi_6^*\,\bar\chi_9^*\,\bar\chi_2^*$.

The lowest free energy solution is a \emph{chiral spin liquid} or \emph{$\Phi\Phi\Phi$-flux phase} \cite{hermele09a,szirmai11b} (or \cite{Corboz13a} for the analog SU(3) system). All of the HS fields have an equal magnitude $|\chi_1|=|\chi_2|=\ldots=|\chi_9|$, and all of the Wilson loops are equal $\Pi_1=\Pi_2=\Pi_3=|\chi_1|^6\,e^{i\Phi}$, where $\Phi=\pm2\pi/3$. As the phase of the Wilson loop is not $0$ or $\pi$, this state is inherently complex. The emerging $\Phi$ phase can be thought as the flux of a fictitious magnetic field that can point up or down perpendicular to the plane of the lattice, accordingly this state is doubly degenerate. The two degenerate states are time reversal partners of each other, therefore, this phase is symmetric to lattice translations and rotations but violates time reversal symmetry --- that causes the chiral nature of this spin liquid state.
The temperature dependence of $|\chi_1|$ together with a sketch of the elementary fluxes is plotted in Fig. \ref{fig:mfsols} a). This shows that the chiral spin liquid state is robust at finite temperature, too, up to a critical temperature which is in the order of the exchange energy $J$.

The next phase, a bit higher in free energy, is the \emph{quasi plaquette} or \emph{$0\pi\pi$-flux phase} \cite{szirmai11b} (or \cite{Corboz13a} for the SU(3) system). In this phase the HS fields inside a central plaquette have higher absolute values than the others: $|\chi_1|=|\chi_2|=\ldots=|\chi_6|>|\chi_7|=|\chi_8|=|\chi_9|$. The flux of the central plaquette is $0$, while the flux of the neighboring plaquettes is $\pi$. This phase is three times degenerate due to the choice of the central plaquette, which can be any of the three independent elementary plaquettes, i.e. the hexagon of the unit cell, or one of its two independent neighbors. This state is time reversal symmetric, but violates the translation symmetry by one hexagon. The temperature dependence of the magnitude of the bigger HS field, say $|\chi_1|$, is plotted in Fig. \ref{fig:mfsols} b) together with an illustration of the arrangement of the fluxes. 

The third phase we have considered is the \emph{plaquette phase} \cite{szirmai11b}. The HS fields in this phase have unit magnitude inside a central plaquette $|\chi_1|=|\chi_2|=\ldots=|\chi_6|$, with a Wilson loop $\Pi_1=1$, and the others are zero $\chi_7=\chi_8=\chi_9=0$. Consequently, $\Pi_2=\Pi_3=0$. This phase is also three times degenerate and violates lattice translation by one plaquette but is invariant under time reversal. This phase can also be characterized by the $|\chi_1|$ as order parameter, whose temperature dependence is shown in Fig. \ref{fig:mfsols}. In this figure the configuration of the plaquette phase is also depicted.

\begin{figure}%[b!]
\begin{center}
  \includegraphics{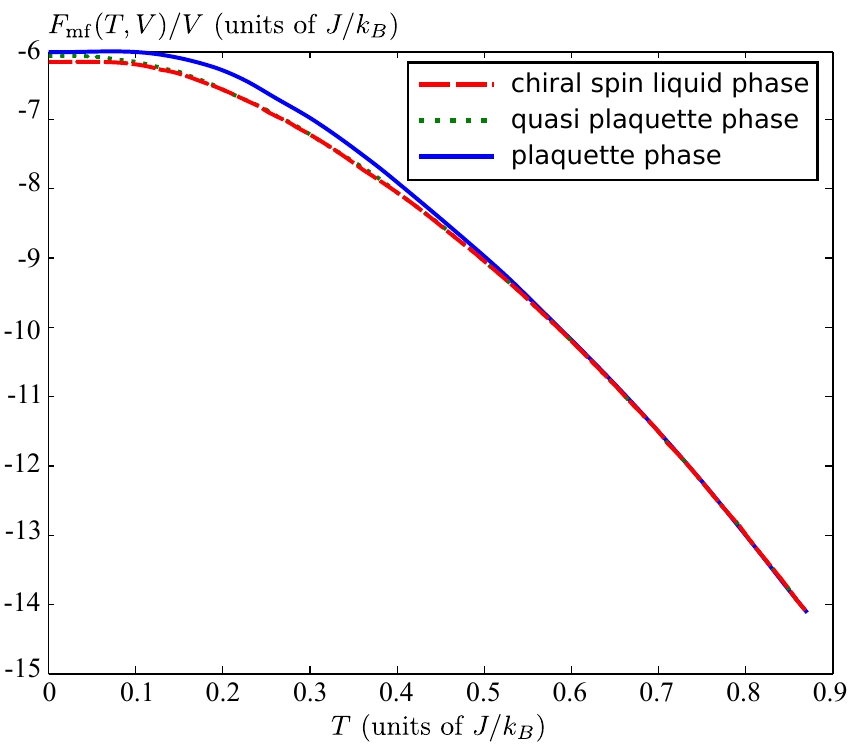}
  \caption{(Color online) The free energy per plaquette $F_{\text{mf}}(T,V)/V$ of the three lowest lying saddle-point solutions.}
  \label{fig:freeenerg}
\end{center}
\end{figure}

The results at zero temperature coincide completely with those found in \cite{szirmai11b}. The interesting thing is that even at finite temperatures these three states remain to be the three lowest free energy saddle-point solutions. For these three lowest lying phases the temperature dependence of the free energy per plaquette is plotted in Fig. \ref{fig:freeenerg}.
We have found that the free energy of the chiral spin liquid state remains slightly below the two other states, however, close to the critical point the free energy curves approach each other with high accuracy. This is in agreement with the behavior of the order parameters. When the temperature is increased the order parameters, the saddle-point values of the HS fields, get smaller and smaller and eventually vanish at a common critical temperature $T_c\approx0.83\,J/k_B$. The order parameters vanish with an exponent of $1/2$, characteristic to the mean-field approximation. Above this temperature the paramagnetic Mott phase is stable, and close to $T_c$ a Ginzburg-Landau type analysis, based on this saddle-point expansion, describes the critical behavior.  

\section{Stability analysis}
\label{sec:stability}

\begin{figure*}
\begin{center}
  \includegraphics{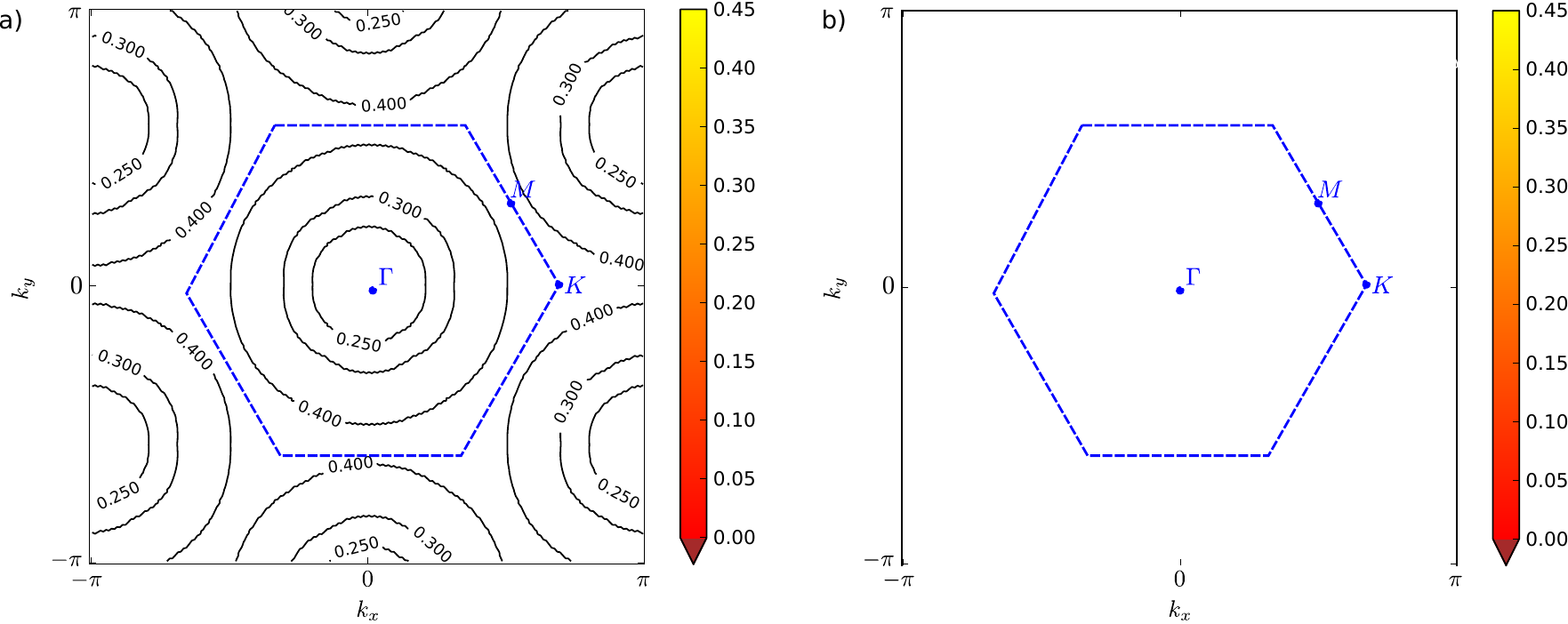}
  \caption{(Color online) The lowest static eigenvalues of the curvature of the effective action in the Brillouin zone at zero temperature for the chiral spin liquid phase a), and for the quasi plaquette phase b).}
  \label{fig:eigenvals}
\end{center}
\end{figure*}
In the framework of the saddle-point approximation the stationary values of the fields are calculated with condition \eqref{eq:SPeqsformal}. Therefore the effective action Eq. \eqref{eq:Seffexp} has no first order contribution in the fluctuations. In order to go beyond the saddle-point approximation we consider the leading correction, which is the second order one in the fluctuations. Thus the functional integral becomes a Gaussian integral, with $S_\text{eff}\approx S_0+S_2+3\,\mathrm{tr}(G_{(0)}\Sigma)^2$. The last two terms, which are quadratic in the fluctuations can be arranged to a convenient matrix form
\begin{subequations}
\label{eqs:curvature}
\begin{multline}
S^{(2)}_{\text{eff}}=\frac{1}{J\beta V}\sum_{i=1}^9\sum_{\hat{q}} \delta\chi^*_i(\hat{q})\delta\chi_i(\hat{q})+3\,\textrm{tr}(G^0\Sigma)^2\\
=\frac{1}{2\beta V}\sum_{\mu,\nu=1}^{24}\sum_{\hat{q}} \phi^*_\mu(\hat{q}) C_{\mu\nu}(\hat{q}) \phi_\nu(\hat{q}).
\end{multline}
The $C_{\mu,\nu}$ kernel is the so called Hessian and provides the curvature of the effective action:
\begin{equation}
\label{eq:Hess}
C_{\mu\nu}(\hat{q})=\frac{\partial^2 S_{\text{eff}}}{\partial\phi_\mu^*(\hat{q})\partial\phi_\nu(\hat{q})}.
\end{equation}
\end{subequations}
The Hessian is a $24\times24$ matrix, and depends on the saddle-point values of the fields $\bar\chi,\bar\varphi$. The derivation and explicit form of $C_{\mu\nu}$ is given in Appendix \ref{appendix}.

Throughout the paper we have assumed that the weight function $e^{-S_\text{eff}}$  in the path integral around the saddle-point configurations takes Gaussian form. In this case the path integral can be evaluated and the partition function is the sum of the Gaussian contributions of the different saddle-points. The first step to test the validity of a specific phase is to check whether the Hessian $C_{\mu\nu}(\hat{q})$ is positive definite at the related configuration \cite{arovas88a}. If $C_{\mu\nu}(\hat{q})$ is positive definite, the weight function in the path integral drops when we move a bit farther from the saddle point, so the saddle-point solution is stable.
The curvature \eqref{eq:Hess} is complex for nonzero Matsubara frequencies, but the sum of the contributions of the $\pm i\nu_m$ pairs always provides a non-negative curvature, since $C_{\mu\nu}(\mathbf{q},-i\nu_m)=C_{\nu\mu}^*(\mathbf{q},i\nu_m)$. Therefore,  it is sufficient to check the stability of the phases for $\nu_m=0$ only.

A minor difficulty still remains. Namely, the scalar potential, introduced in Eq. \eqref{eq:deltarep} has to be purely imaginary for the representation of the Dirac delta functions. However, the saddle-point equations provide real solutions for $\bar\varphi_s$. Such solutions are physical, and one can interpret the homogeneous and real $\bar\varphi$ as the chemical potential of the system. Consequently the functional integral representation \eqref{eq:deltarep} has to be understood after an analytical continuation $\delta\varphi_s\rightarrow i\delta\varphi_s$. Hence in the Hessian the curvature is changed from positive to negative along the 6 directions of the scalar potential. The easiest way to treat the problem is to perform the Gaussian integral over $\delta\varphi_s$, as was done e.g. in Ref \cite{marston89a} to arrive to an effective action only for the $\delta\chi$ fields,
\begin{equation}
\label{eq:Hesseff}
S^{(2)'}_{\text{eff}}=\frac{1}{2\beta V}\sum_{k,l=1}^{18}\sum_{\hat{q}}
\phi_k^*(\hat q)
\widetilde C_{kl}(\hat{q})
\phi_l(\hat q),
\end{equation}
with $\phi_k(\hat q)$ a vector of 18 elements, obtained from $\phi_\mu(\hat q)$ by simply dropping the last 6 entries. $\widetilde C_{kl}(\hat{q})$ is a $18\times18$ matrix whose elements are formed from the matrix $C_{\mu\nu}(\hat{q})$. Its final form is also given in Appendix \ref{appendix}.

Now we reduced the stability problem to the eigenvalue analysis of the $\tilde{C}_{kl}$ matrix: if all the 18 eigenvalues of $\tilde{C}_{kl}$ for each $\mathbf{q}$ (i.e. in the whole Brillouin zone) are positive, $\tilde{C}_{kl}$ is positive definite, the path integral remains Gaussian, so the saddle-point approximation is reliable.

We have determined the spectrum of $\tilde{C}_{kl}$ for each of the three low lying states at low temperatures. In Fig. \ref{fig:eigenvals} we plot the lowest nonzero eigenvalues for the stability matrix $\widetilde C_{ij}(\mathbf{q},0)$ in the Brillouin zone for the chiral spin liquid phase a), and for the quasi plaquette phase b). The chiral spin liquid phase is stable against perturbations, as the lowest nonzero eigenvalue is positive everywhere. Contrary, the lowest nonzero eigenvalue of the curvature of effective action in case of the quasi plaquette phase develops prominent negative values around the $\Gamma$ point. This phase turns out to be unstable. Note that among the 18 eigenvalues (for every $\mathbf{q}$) of the stability matrix $\widetilde C_{ij}(\mathbf{q},0)$ we have some flat zero modes corresponding to the local gauge symmetry. Both the chiral spin liquid and quasi plaquette phases have 6 such flat bands, since 6 of the link variables can be chosen real with the help of gauge fixing. 

In the plaquette phase all eigenvalues are flat, because the lattice is formed by the completely disjoint plaquettes and no momentum dependence remains. 
In this phase three of the links are zero, and we can fix only 5 of the 6 remaining $\bar\chi$ links to be real. Correspondingly only 5 flat zero modes remain.  All the other (nonzero) eigenvalues are positive, which means that the plaquette phase is stable. 

It is worth to emphasize that, however, the free energy analysis suggests a strong competition of the quasi plaquette state and the chiral SL state (see Fig. \ref{fig:freeenerg}), the stability analysis shows that the quasi plaquette state collapses towards the lower free energy solution. The situation is similar to the case when the $\pi$-flux state of the SU(2) system on square lattice turns out to be unstable and collapse into the so called "box" state \cite{marston89a,dombre89a}. Accordingly, only the chiral SL state remains the lowest lying state at least up to $k_\mathrm{B} T \sim 0.5 J$, where its free energy starts to compare with that of the plaquette state. Above this temperature the two states have practically the same free energy, and with simple cooling it can not be predicted which phase will stabilize. Nevertheless, the two phases have different symmetries and most importantly different topological properties that may allow to select the demanded state. For example during cooling and yet in the high temperature phase via imprinting a synthetic external gauge field to generate the nontrivial topology of the CSL state. At the low temperature state the enforced topological property remains even when the external constraints are switched off.

\section{Structure factor}
\label{sec:structfact}

\begin{figure*}
\begin{center}
  \includegraphics{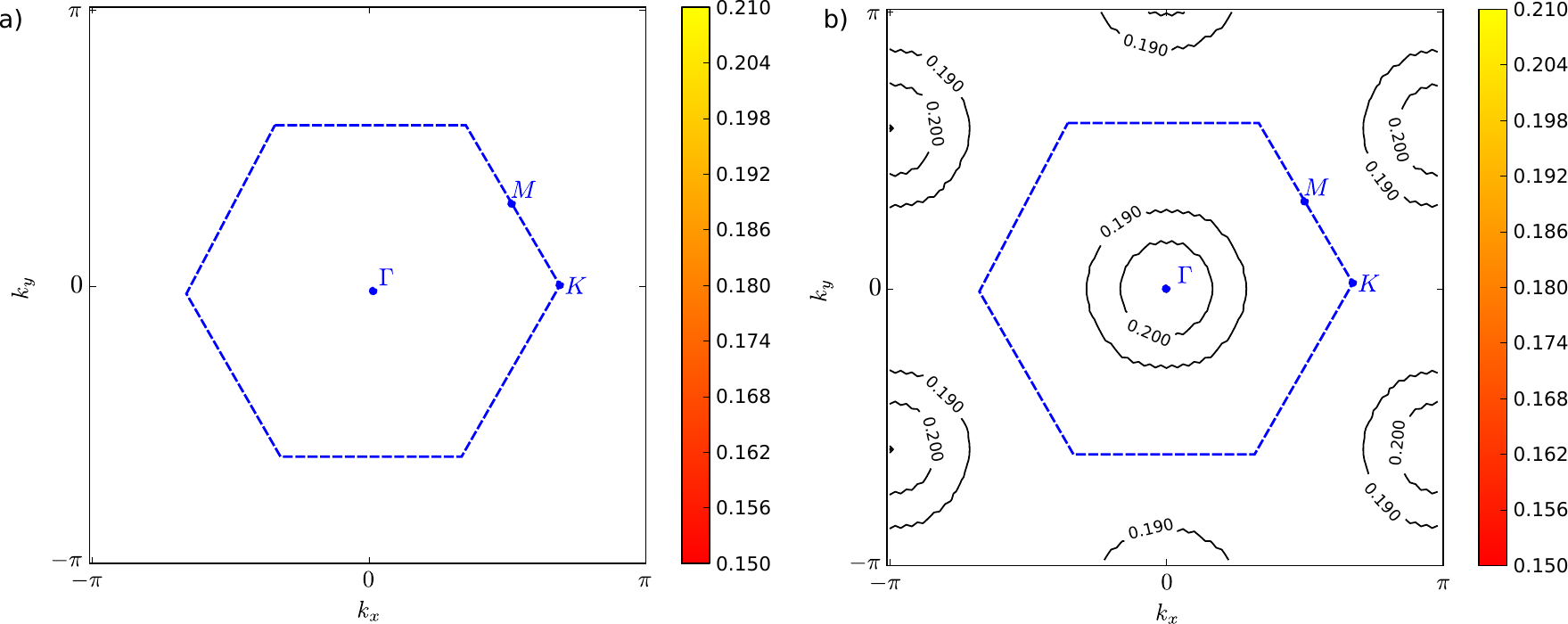}
  \caption{(Color online) The static structure factor $\mathcal{S}(\mathbf{q},0)$ in the Brillouin zone at zero temperature for the chiral spin liquid phase a), and for the quasi plaquette phase b).}
  \label{fig:staticstructfact}
\end{center}
\end{figure*}

In order to study experimental signatures of the different spin liquid and valence bond phases one can look for the experimentally measurable spin-spin correlation function, also known as the structure factor:
\begin{equation}
\mathcal{S}(\mathbf{r},\tau,\mathbf{r}',0) = \left< \left(S^z(\mathbf{r},\tau) S^z(\mathbf{r'},0) \right) \right>,
\end{equation}
where $S^z(\mathbf{r}_{mns}) = \sum_{\alpha\beta}F^z_{\alpha \beta}c^\dagger_{s\alpha}(\mathbf{r}_{mn}) c_{s\beta}(\mathbf{r}_{mn})$ is the $z$-component of the spin operator of a spin-5/2 alkaline earth atom at site $\mathbf{r}_{mns}$, accordingly, $F^z=\mathrm{diag}(5/2,3/2,1/2,-1/2,-3/2,-5/2)$ is the $z$-component of the three SU(2) generators in 6 dimensional representation. In momentum space and Matsubara representation
\begin{multline}
\mathcal{S}_{s's}(\mathbf{q},i\nu_m) = \frac{35}{2}\frac{1}{V \beta}\sum_{\mathbf{k},a,b} \frac{n(\varepsilon^{(a)}_\mathbf{k})-n(\, \varepsilon^{(b)}_{\mathbf{k}+\mathbf{q}} \, )}{i \nu_m + \varepsilon^{(a)}_{\mathbf{k}}-\varepsilon^{(b)}_{\mathbf{k}+\mathbf{q}}}\\
\times v^{(a)*}_{s}(\mathbf{k}) v^{(a)}_{s'}(\mathbf{k}) v^{(b)*}_{s'}(\mathbf{k}+\mathbf{q})v^{(b)}_{s}(\mathbf{k}+\mathbf{q}),
\label{spin_struct_psinko}
\end{multline}
where the numeric factor comes from spin summation: $\sum_{\alpha\beta} F^z_{\alpha\beta} F^z_{\beta\alpha} = 35/2$, and the fermionic occupation number at finite temperature is given by the Fermi distribution function $n(\varepsilon) = [e^{\beta \varepsilon} + 1]^{-1}$. $\varepsilon_{\mathbf{k}}$ and $v_{s}(\mathbf{k})$ are the eigenenergies and eigenvectors of $H^{(0)}$, respectively, as they were introduced in Eq. \eqref{eq:H0diag}. The structure factor  $\mathcal{S}_{ss'}(\mathbf{q},i\nu_m)$ in Eq. \eqref{spin_struct_psinko} is a $6\times 6$ matrix in the sublattice space. In order to take into account the total contribution of the unit cell, one needs to consider its trace, $\mathcal{S}(\mathbf{q},i\nu_m)=\sum_s \mathcal{S}_{ss}(\mathbf{q},i\nu_m)$.
We plot the static structure factors in Fig. \ref{fig:staticstructfact} of the chiral spin liquid phase a) and of the quasi plaquette phase b). That of the plaquette phase is completely flat due to its dispersionless spectra, and is not shown. For the other two low lying saddle-point solutions the static structure factors look completely different, both carry unambiguous features to identify them. In the chiral spin liquid phase the structure factor has a minimum at the center of the Brillouin zone (the $\Gamma$ point) and it has maxima at the $K$ points. In contrary, in the quasi plaquette phase the structure factor is peaked close to the $\Gamma$ point and has minima around the edge of the Brillouin zone. Since the experimentally measurable structure factor of the three lowest lying saddle-point configurations show completely different behavior, it is a suitable tool to distinguish between them. 

\begin{figure}[b!]
\begin{center}
  \includegraphics{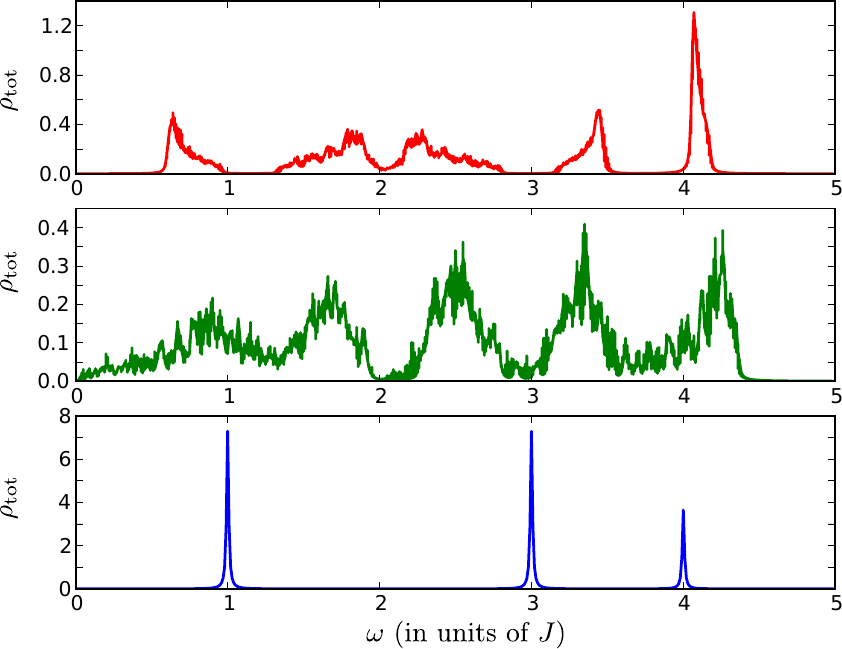}
  \caption{(Color online) The spectral density $\rho_\text{tot}(\omega)$ at zero temperature of the chiral spin liquid phase a), the quasi plaquette phase b), and the plaquette phase c).}
  \label{fig:specdens}
\end{center}
\end{figure}

By radio-frequency-spectroscopy one can measure the spectral density integrated to the whole lattice. This quantity can also be extracted from the structure factor given by Eq. \eqref{spin_struct_psinko}. 
It can be obtained by analytically continuing $S(\mathbf{q},i\nu_m)$ on the the upper half plane of complex frequencies and through the real axis by setting $i\nu_m\rightarrow\omega+i\eta$.
\begin{equation}
\label{eq:spectraldensity}
\rho_\text{tot}(\omega)=\sum_{\mathbf{q}}\mathrm{Im}\,\mathcal{S}(\mathbf{q},\omega+i\eta),
\end{equation}
with $\eta$ being an infinitesimally small number. In our calculation we set its value to $10^{-2}$ (in units of the coupling strength $J$). We plot the spectral density of the three mean-field solutions in Fig. \ref{fig:specdens}. The spectral function of the three different phases looks completely different indicating some characteristic features of the specific phase. While the chiral spin liquid phase (a), and the plaquette phase (c) are gapped phases, there are no accessible states up to a gap, the quasi plaquette phase (b) is gapless. Furthermore, the plaquette phase is very simple, it contains only 3 delta peaks, also because of the lack of dispersion of the fermion energies. The first two peaks have twice the strength of the third one according to the fermion spectrum that consists of 4 dispersionless flat bands. From these 4 flat bands the lowest and the highest energy bands have only one state for every $\mathbf{q}$ momentum, and between them two doubly degenerate flat bands can be found. At zero temperature the lowest band is occupied and the higher bands are empty. By exciting a fermion to the upper bands the middle two have twice the number of states than the last one.

\section{Summary and conclusions}
\label{sec:summary}

In this paper we have studied finite temperature spin liquid and valence bond solid phases of alkaline earth atoms on a honeycomb lattice. We have determined the lowest free energy solutions in a mean-field level and studied their temperature dependence. We have found that the spin liquid and valence bond solutions melt at a common critical temperature in the order of the superexchange interaction. We also studied the stability of the encountered states and have shown that the chiral spin liquid state is indeed the stable saddle-point and the quasi plaquette state is unstable against fluctuations with zero momentum. The plaquette state is also stable, though with higher free energy. 

In experiments cooling the fermions towards quantum degeneracy is a difficult to achieve goal. Though systems of fermions with SU(6) symmetry seem to be more complicated than the two component SU(2) fermions, it turned out recently, that  experimentally it is easier to cool them towards the magnetic transition \cite{taie12a,cai13a} because an isolated 6 component atom can carry away much more entropy than a 2 component one. This effect is similar to the Pomeranchuk cooling first observed in solid $^3$He \cite{richardson97a}. Combining the Pomeranchuk cooling with lattice shaking \cite{hauke12a}, which can imprint a nontrivial topology to the system, it might be possible to directly cool the SU(6) symmetric Mott insulator into the topologically nontrivial CSL state even if its free energy is close to other VBS like phases.  

We have also studied the experimentally measurable signatures of the mean-field states, namely the spin-spin correlation function, and its spectral function, of the alkaline earth atoms. We have shown that these quantities qualitatively differ for the different mean-field solutions and therefore can be used as a smoking gun in experiments to reveal the realized specific phase.

A further method to unambiguously verify whether the state is the topologically nontrivial CSL state would be by probing the existence of chiral edge states. In Ref. \cite{goldman13a} a simple method was introduced to image directly the chiral edge modes. 
The method is based on the shaping of the atomic cloud with an extremely steep confining potential and then suddenly removing the confining wall. After the sudden release of the gas the spatial density evolution of the bulk as well as the edge modes are traceable, and the movement of the chiral edge states becomes directly visible. In order to get better visibility on longer time scale, there are two important requirements. On one hand, a large initial occupancy of the edge states is required, and on the other hand, it is also important that the edge states contribution to the density
has to remain spatially separated from the bulk during the evolution. The former requirement can be tuned via the density (Fermi energy), while the latter one can be ensured by populating a quasi-dispersionless bulk band with non-zero Chern number. In Ref. \cite{szirmai11b} we demonstrated that the chiral spin liquid state lowest lying bulk band is almost flat, and completely filled, therefore, it is expected that the emerging topological edge states of the chiral SL state as unequivocal signatures of its topological nature can be detected. 

\section*{Acknowledgements}

We acknowledge enlightening discussions with L. Fallani, M. Lajk\'o, F. Mila, and K. Penc. This work is supported by the Hungarian Academy of Sciences (Lend\"ulet Program, LP2011-016), NORT (ERC\_HU\_09 OPTOMECH), the NRF (OTKA PD104652, K100908), EU (SIQS), ERC (QUAGATUA), Spanish MINCIN (FIS2008-00784 TOQATA), Generalitat de Catalunya (2009-SGR1289). G.Sz. also acknowledges support from the J\'anos Bolyai Scholarship. ML also acknowledges support from Alexander von Humboldt Stiftung.

\appendix
\section{Feynman rules}
\label{appendix}

\begin{figure}[tb]
\begin{center}
  \includegraphics{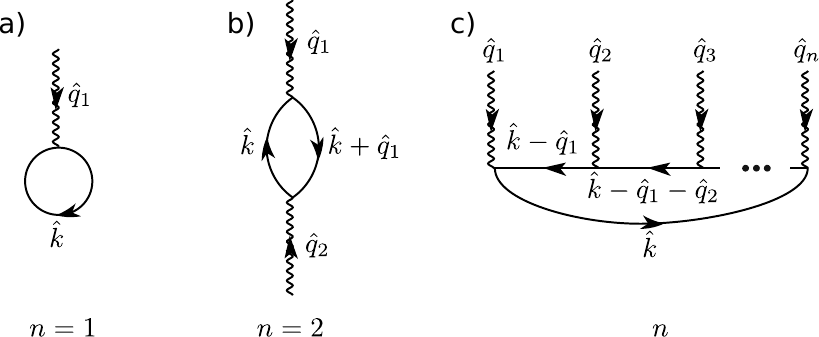}
  \caption{Feynman graphs representing the expansion Eq. \eqref{log_term_appendix_alex}. We have plotted the first order a), the second order b) and the nth order c) diagrams. The straight lines represent the free fermion propagator, while the wiggly line is for the $\Sigma$ vertex. The sum of the incoming momentum has to be zero.}
  \label{fig:Feynman_graphs}
\end{center}
\end{figure}

In this appendix our goal is to derive
the curvature of the effective action and give the explicit form of the stability matrix \eqref{eq:Hess} and the one in Eq. \eqref{eq:Hesseff}. To this end we need to evaluate
\begin{multline}
\mathrm{tr}(G_0\Sigma)^n
=\sum_{\hat{k},\hat{q}_1\ldots\hat{q}_n} \mathrm{Tr}\Big[G_{(0)}(\hat{k})\Sigma(\hat{k}-\hat{q}_1,\hat{q}_1)\\
\times G_{(0)}(\hat{k}-\hat{q}_1)\Sigma(\hat{k}-\hat{q}_1-\hat{q}_2,\hat{q}_2)\times \ldots \\
\times G_{(0)}(\hat{k}-\hat{q}_1-\ldots-\hat{q}_{n-1})\Sigma(\hat{k}-\hat{q}_1-\ldots-\hat{q}_{n},\hat{q}_n)\Big],
\label{log_term_appendix_alex}
\end{multline}
where $\sum_n\hat{q}_n=0$. In the first line $\mathrm{tr}$ is a sum for momentum, Matsubara frequency and sublattice index. In the second line $\mathrm{Tr}$ is understood only in the sublattice indices as the sum is explicitly indicated for the momentum and Matsubara frequencies. Eq. \eqref{log_term_appendix_alex} is prone to be represented by Feynman diagrams. At a given order (say $n$) we have exactly $n$ free fermion propagators $G_{(0)}$, represented by straight lines, and also $n$ incoming vertices $\Sigma$, represented by wiggly lines. The arrows show the direction of the transfer of momentum. The entire graph is connected and contains a single loop with momentum and Matsubara frequency conservation. For illustration we have shown the first (a), the second (b), and the general, nth order (c) graphs in Fig. \ref{fig:Feynman_graphs}. Note that both $G_{(0)}$ and $\Sigma$ are matrices in the sublattice index.

For the curvature of the effective action \eqref{eqs:curvature} we need to evaluate
\allowdisplaybreaks
\begin{multline}
\label{eq:secondordertotrace}
R_{\mu\nu}(\hat{q})\equiv\mathrm{tr}\left[\frac{\partial^2 \left(G_{(0)}\Sigma\right)^2}{\partial \phi_\mu^*(\hat{q})\partial \phi_\nu(\hat{q})}\right]\\
=\mathrm{tr}\left(G_{(0)}(\hat{k})
\frac{\partial \Sigma(\hat{k}+\hat{q},-\hat{q})}{\partial \phi_\mu^*(\hat{q})}
G_{(0)}(\hat{k}+\hat{q})
\frac{\partial \Sigma(\hat{k},\hat{q})}{\partial \phi_\nu(\hat{q})}
\right)\\
=\sum_{\stackrel{\mathbf{k},i\omega_n}{s_1,s_2,s_3,s_4}}G_{(0)s_1,s_2}(\mathbf{k},i\omega_n)G_{(0)s_3,s_4}(\mathbf{k}+\mathbf{q},i\omega_n+i\nu_m)\\
\times\frac{\partial \Sigma_{s_2,s_3}(\hat{k}+\hat{q},-\hat{q})}{\partial \phi_\mu^*(\hat{q})}
\frac{\partial\Sigma_{s_4,s_1}(\hat{k},\hat{q})}{\partial \phi_\nu(\hat{q})}\\
=\sum_{\stackrel{\mathbf{k},a,b}{s_1,s_2,s_3,s_4}}\frac{v^{(a)}_{s_1}(\mathbf{k})\,v^{(a)*}_{s_2}(\mathbf{k})\,v^{(b)}_{s_3}(\mathbf{k}+\mathbf{q})\,v^{(b)*}_{s_4}(\mathbf{k}+\mathbf{q})}{i\nu_m+\varepsilon^{(a)}_\mathbf{k}-\varepsilon^{(b)}_{\mathbf{k}+\mathbf{q}}}\\
\times\left[n(\varepsilon^{(a)}_\mathbf{k})-n(\varepsilon^{(b)}_{\mathbf{k}+\mathbf{q}})\right]
\frac{\partial \Sigma_{s_2,s_3}(\hat{k}+\hat{q},-\hat{q})}{\partial \phi_\mu^*(\hat{q})}
\frac{\partial\Sigma_{s_4,s_1}(\hat{k},\hat{q})}{\partial \phi_\nu(\hat{q})}.
\end{multline}
The derivatives of the self-energies are again easily evaluated with the help of Eq. \eqref{eq:sigma}. Combining Eq. \eqref{eq:secondordertotrace} with Eqs. \eqref{eqs:curvature} we arrive to the $24\times24$ Hessian matrix
\begin{equation}
C_{\mu\nu}(\hat q)=6\,R_{\mu\nu}(\hat q)+\frac{1}{J}\sum_{i=1}^{18}\delta_{\mu,i}\delta_{\nu,i}.
\end{equation}
With the help of Eq. \eqref{eq:secondordertotrace} it can be directly checked that
\begin{equation}
\label{eq:Csymm}
C_{\mu\nu}(\textbf{q},i\nu_m)= C_{\nu\mu}^*(\textbf{q},-i\nu_m).
\end{equation}

Finally let us construct the Hessian matrix $\widetilde C_{kl}(\hat q)$ appearing in Eq. \eqref{eq:Hesseff} after integrating out the $\delta\varphi$ fields. For a convenient notation let us introduce submatrices of the original $24\times24$ matrix $C_{\mu\nu}(\hat q)$, such that  
\begin{widetext}
\begin{equation}
\label{Cmatrix}
C_{\mu\nu}(\hat q)=
\left[
\begin{array}{c c c c c c c c}
C_{1,1}(\hat q) & C_{1,2}(\hat q) & \dots & C_{1,18}(\hat q) & W_{1,1}(\hat q)   & W_{1,2}(\hat q) & \dots    & W_{1,6}(\hat q) \\
 &  \ddots  &    &   \vdots   & \vdots    &       &       &  \vdots   \\
 \\
 &       &  & C_{18,18}(\hat q)  & W_{18,1}(\hat q)    &     \dots      &     & W_{18,6}(\hat q) \\
 &  &  &  & E_{1,1}(\hat q)   &                        & \dots    & E_{1,6}(\hat q) \\
 &  &  &  & &       \ddots     &               & \vdots                    \\
 &  &  &  &   &   &  & E_{6,6}(\hat q)
\end{array}\right]
\end{equation}
\end{widetext}
where $C_{kl}$ is a $18\times18$, $W_{ks}$ is a $18\times6$ and $E_{sr}$ is a $6\times6$ matrix. The elements below the diagonal are understood to be filled according to the relation \eqref{eq:Csymm}. With the help of this notation, the path integral over the $\delta\varphi$ fields, in the Gaussian approximation Eq. \eqref{eqs:curvature} is performed by
\begin{equation}
\int D[\delta\varphi]\,e^{-S^{(2)}_\text{eff}[\delta\chi,\delta\chi^*,\delta\varphi]}=\frac{1}{\sqrt{\det E}}e^{-S^{(2)'}_\text{eff}[\delta\chi,\delta\chi^*]},
\end{equation}
with $S^{(2)'}_\text{eff}$ given in Eq. \eqref{eq:Hesseff} with the matrix
\begin{multline}
\widetilde C_{kl}(\mathbf{q},i\nu_m)=C_{kl}(\mathbf{q},i\nu_m)\\
-\sum_{r,s=1}^6 W_{ks}(\mathbf{q},i\nu_m)E^{-1}_{sr}(\mathbf{q},i\nu_m)W_{lr}^*(\mathbf{q},-i\nu_m).
\end{multline}

\end{document}